\documentclass[a4paper,12pt]{article}
\usepackage{graphics}

\begin{document}

\title{Coherent and incoherent pair creation by a photon
in oriented single crystal}
\author{V. N. Baier
and V. M. Katkov\\
Budker Institute of Nuclear Physics,\\ Novosibirsk, 630090,
Russia}

\maketitle

\begin{abstract}
The new approach is developed for study of 
electron-positron pair production by a photon in 
oriented single crystal. It permits indivisible
consideration of both coherent and incoherent mechanisms
of pair creation and includes the action of field of
axis (or plane) as well as the multiple scattering of
particles of the created pair (the Landau-Pomeranchuk-Migdal 
(LPM) effect). From
obtained integral probability of pair creation, it follows
that multiple scattering appears only for relatively
low energy of photon, while at higher photon energy the field
action excludes the LPM effect. The found probabilities
agree quite satisfactory with recent CERN experiment.

\end{abstract}

\newpage

It is known that mechanism of pair production by a photon in
oriented single crystal differs substantially from the mechanism
of independent pair creation at separate centers acting in an
amorphous medium (the Bethe-Heitler mechanism). In crystal the
coherent interaction of a photon with many centers occurs. Under
some generic assumptions the general theory of the coherent pair
creation mechanism was developed in \cite{BKS1}. 
If the photon angle of incidence
$\vartheta_0$ (the angle between photon momentum {\bf k} and the
axis (or plane)) is small $\vartheta_0 \ll V_0/m$, where $V_0$ is
the characteristic scale of the potential, the field $E$ of the
axis (or plane) can be considered constant over the pair formation
length and the constant-field approximation is valid. In this case
the behavior of pair production probability is determined by the
parameter
\begin{equation}
\kappa=\frac{\omega}{m}\frac{E}{E_0},
\label{1}
\end{equation}
where $\omega$ is the photon energy, $m$ is the electron mass,
$E_0=m^2/e=1.32 \cdot 10^{16}$ V/cm is the critical field, the
system $\hbar=c=1$ is used. The very important feature of
coherent pair creation mechanism is the strong enhancement of its
probability at high energies (from factor $\sim 10$ for main 
axes in crystals of
heavy elements like tungsten to factor $\sim 160$ for diamond)
comparing with the Bethe-Heitler mechanism. If $\vartheta_0 \gg
V_0/m$ the theory passes over to the coherent pair production 
theory (see e.g. \cite{BKS}, Sec.13.1). Side by side with 
coherent mechanism the incoherent
mechanism of pair creation is acting. In oriented crystal this
mechanism changes also with respect to an amorphous medium
\cite{BKS2}. The details of theory and description of experimental
study of pair production which confirms the mentioned enhancement
can be found in \cite{BKS}.  The item continues to attract
attention \cite{N} and new experiments are performed recently
\cite{KKM}, \cite{B}. In high-energy region the multiple scattering
of particles of created pair (the Landau-Pomeranchuk-Migdal (LPM)
effect) is influenced in probability of pair creation. This paper
is devoted to study of the problem using the new theoretical
approach..

The properties of pair creation process are connected directly
with details of motion of particles of created pair. The momentum
transfer from a particle to a crystal we present in a form
${\bf q}=<{\bf q}>+{\bf q}_s$, where $<{\bf q}>$ is the mean value
of momentum transfer calculated with averaging over thermal(zero)
vibrations of atoms in a crystal. The motion of particles in an
averaged potential of crystal, which corresponds to the momentum
transfer $<{\bf q}>$, determines the coherent mechanism of pair
creation. The term ${\bf q}_s$ is attributed to the random
collisions of particle which define the incoherent mechanism of
discussed process. Such random collisions we will call
"scattering" since $<{\bf q}_s>=0$. If the formation length of
pair creation process is large with respect to distances between
atoms forming the axis, the additional averaging over the atom
position should be performed.

Here we consider case $\vartheta_0 \ll V_0/m$. Than the distance
from axis $\mbox{\boldmath$\varrho$}$ as well as the transverse
field of the axis can be considered as constant over the
formation length. The process of the electron-positron
pair creation in this case is one of interesting applications of the
theory of propagation of a photon in a medium in presence of an
external field \cite{BK}. In the problem under consideration we
have both the dense matter with strong multiple scattering and
high field of crystal axis.

For an axial orientation of crystal the ratio of the atom density
$n(\varrho)$ in the vicinity of an axis to the mean atom density 
$n_a$ is
\begin{equation}
\frac{n(x)}{n_a}=\xi(x)=\frac{x_0}{\eta_1}e^{-x/\eta_1},
\label{2}
\end{equation}
where
\begin{equation}
x_0=\frac{1}{d n_a a_s^2}, \quad  \eta_1=\frac{2
u_1^2}{a_s^2},\quad x=\frac{\varrho^2}{a_s^2},
\label{3}
\end{equation}
Here $\varrho$ is the distance from axis, $u_1$ is the amplitude
of thermal vibration, $d$ is the mean distance between atoms
forming the axis, $a_s$ is the effective screening radius of the
axis potential (see Eq.(9.13) in \cite{BKS})
\begin{equation}
U(x)=V_0\left[\ln\left(1+\frac{1}{x+\eta} \right)-
\ln\left(1+\frac{1}{x_0+\eta} \right) \right].
\label{4}
\end{equation}
The local value of parameter $\kappa(x)$ (see Eq.(\ref{1})) which
determines the probability of pair creation in the field Eq.(\ref{4} is
\begin{equation}
\kappa(x)=-\frac{dU(\varrho)}{d\varrho}\frac{\omega}{m^3}=\kappa_s
\frac{2\sqrt{x}}{(x+\eta)(x+\eta+1)},\quad
\kappa_s=\frac{V_0 \omega}{m^3a_s}\equiv \frac{\omega}{\omega_s}.
 \label{5}
\end{equation}
The parameters of the axial potential for the ordinarily used
crystals are given in Table 9.1 in \cite{BKS}.
The particular calculation below will be done for tungsten 
crystals studied in \cite{KKM}. The relevant
parameters are given in Table 1.

In an amorphous medium (or in crystal in the case of random
orientation) the LPM effect becomes essential for heavy elements
at the characteristic photon energy $\omega_e \sim 10$~TeV
\cite{BK1} and this value is inversely proportional to the
density. In the aligned case the ratio $\xi(0)$ Eq.(\ref{2}) may
attain the magnitude $\xi(0) \sim 10^3$ in cold crystals. So the
characteristic photon energy becomes $\omega_0 =
\omega_e/\xi(0) \sim 10$~GeV. It is useful to compare the
characteristic energy $\omega_0$ with known threshold energy
$\omega_t$ for which the probability of pair creation in the axis
field becomes equal to the Bethe-Maximon probability, see
Sec.12.2 and Table 12.1 in \cite{BKS}. For small value of the
parameter $\kappa$ the probability of coherent pair creation is
(see Eq.(12.11) in \cite{BKS})
\begin{equation}
W^F=\frac{9}{32}\sqrt{\frac{\pi}{2}}\frac{\alpha m^2}{\omega x_0}
\frac{\kappa_m^2}{\sqrt{-\kappa''_m}}\exp(-8/3\kappa_m).
\label{5a}
\end{equation}
Here $\kappa_m$ is the maximal value of the parameter $\kappa(x)$
Eq.(\ref{5}) (which defines the value of $\omega_t$)
\begin{equation}
\kappa_m=\kappa(x_m),\quad x_m=\frac{1}{6}
\left(\sqrt{1+16\eta(1+\eta)}-1-2\eta \right),\quad
\kappa''_m=\kappa''(x_m)
\label{6}
\end{equation}
We present it in the form $\kappa_m=\omega/\omega_m$. Than we find
that $\omega_t \sim \omega_m \sim \omega_0$ for main axes of
crystals of heavy elements. So at $\omega \sim \omega_t$ all the
discussed effects are simultaneously essential in these crystals.
In crystals of elements with intermediate $Z$ (Ge, Si, diamond)
the ratio $\omega_t/\omega_m \sim 1$ but $\omega_m/\omega_0 \ll 1$. 
So, one can neglect the LPM effect at
$\omega \sim \omega_t$.

At $\omega \ll \omega_t$ the pair creation integral cross
section for incoherent mechanism in oriented crystal has the
form (see Eq.(26.30) in \cite{BKS})
\begin{eqnarray}
&&\sigma_p=\frac{28Z^2\alpha^3}{9m^2}\left[L_0-\frac{1}{42}
-h\left(\frac{u_1^2}{a^2} \right)\right],\quad L_0=\ln(ma)+
\frac{1}{2}-f(Z\alpha),
\nonumber \\
&& h(z)=-\frac{1}{2}\left[1+(1+z)e^{z}{\rm Ei}(-z) \right],\quad
a=\frac{111Z^{-1/3}}{m},
\nonumber \\
&& f(\xi)={\rm Re}
\left[\psi(i+i\xi)-\psi(1) \right]=\sum_{n=1}^{\infty}
\frac{\xi^2}{n(n^2+\xi^2)},
\label{7}
\end{eqnarray}
where $\psi(z)$ is the logarithmic derivative of the gamma
function, Ei($z$) is the integral exponential function,
$f(\xi)$ is the Coulomb correction.
The cross section differs from the Bethe-Maximon cross
section only by the term $h(u_1^2/a^2)$ which reflects the
nongomogeneity of atom distribution in crystal. For $u_1\ll a$
one has
$h(u_1^2/a^2) \simeq -(1+C)/2+\ln(a/u_1),~C=0.577..$ and so 
this term characterizes the new value of upper
boundary of impact parameters $u_1$ contributing to the
value $<{\bf q}_s^2>$ instead of screening radius $a$
in an amorphous medium.

The influence of axis field on the incoherent pair creation process
begins when $\omega$ becomes close to $\omega_m$. For small values
of the parameter $\kappa_m$ the correction to the cross section
Eq.(\ref{6}) is (see Eq.(7.137) in \cite{BKS})
\begin{eqnarray}
&& \Delta \sigma_{p}=\frac{176}{175}\frac{Z^2\alpha^3}{m^2}
\overline{\kappa^2} \left(L_u-\frac{1789}{1980}\right),
\nonumber \\
&& \overline{\kappa^2}=\int_{0}^{\infty}\frac{dx}{\eta_1}
e^{-x/\eta_1}\kappa^2(x),\quad
L_u=L_0-h\left(\frac{u_1^2}{a^2}\right).
 \label{8}
\end{eqnarray}

The coherent and incoherent contribution to pair creation can
separated also for $\kappa_m \gg 1~(\omega \gg \omega_m)$. In
this case one can use the perturbation theory in calculation of 
the probability of incoherent process and neglect the LPM effect 
because of domination of the
coherent contribution and additional suppression (by the axis
field) the incoherent process. In this case the local cross
section of pair creation has the form (see Eq.(7.138) in
\cite{BKS})
\begin{equation}
\sigma_p(x)=\frac{8Z^2\alpha^3 \Gamma^3(1/3)}{25
m^2(3\kappa(x))^{2/3}\Gamma(2/3)}
\left(L_u+0.4416+\frac{1}{3}\ln\kappa(x)\right).
\label{9}
\end{equation}
Averaging the function $(\kappa(x))^{-2/3}$ and
$\ln\kappa(x)(\kappa(x))^{-2/3}$ over $x$ according with
Eq.(\ref{8}) one can find the effective value of upper boundary
of the transverse momentum transfer ($\propto m\kappa^{1/3}_m$
instead of $m$) which contributes to the value $<{\bf q}_s^2>$.
Using the obtained results we determine the effective logarithm
$L$ by means of interpolation procedure
\begin{equation}
L=L_0g,\quad
g=1+\frac{1}{L_0}\left[-\frac{1}{42}-h
\left(\frac{u_1^2}{a^2}\right)
+\frac{1}{3}\ln\left
(\frac{6-3\kappa^2_m+3\kappa^3_m}{6+\kappa^2_m}\right) \right].
 \label{10}
\end{equation}

Let us introduce the local characteristic energy
\begin{equation}
\omega_c(x)=\frac{m}{4\pi Z^2\alpha^2\lambda_c^3n(x)L}=
\frac{\omega_e(n_a)}{\xi(x)g}=\frac{\omega_0}{g}e^{x/\eta_1},
\label{11}
\end{equation}
where $\lambda_c=1/m$. In this notations
the local probability for small values of $\kappa_m$ 
and $\omega/\omega_0$ has a form (see Eq.(7.137) in
\cite{BKS} and Eq.(2.23) in \cite{BK1})
\begin{equation}
W(x)=\frac{7}{9\pi}\frac{\alpha m^2}{\omega_c(x)}
\left[1+\frac{396}{1225}\kappa^2(x)-\frac{3312}{2401}
\frac{\omega^2}{\omega^2_c(x)} \right], 
\label{12}
\end{equation}
where the term with $\kappa^2(x)$ arises due to the field action and
the term with $\omega^2/\omega^2_c(x)$ reflects influence of
multiple scattering (the LPM effect).
Averaging this expression over $x$ we have
\begin{eqnarray}
&& \int_0^{\infty}\frac{dx}{x_0}\frac{1}{\omega_c(x)}=
\frac{g}{\omega_0}\frac{\eta_1}{x_0}=\frac{g}{\omega_e(n_a)},\quad
\int_0^{\infty}\frac{dx}{x_0}\frac{1}{\omega^3_c(x)}=
\frac{g}{\omega_e(n_a)}\frac{g^2}{3\omega_0^2},
\nonumber \\
&& \int_0^{\infty}\frac{dx}{x_0}\frac{\kappa^2(x)}{\omega_c(x)}=
\frac{g}{\omega_e(n_a)}\overline{\kappa^2},
\nonumber \\
&& W \equiv \overline{W(x)}=W_0 g
\left[1+\frac{396}{1225}\overline{\kappa^2}
-\frac{1104}{2401}\left(\frac{\omega g}{\omega_0} \right)^2 \right], 
 \label{13}
\end{eqnarray}
where $W_0$ is 
\begin{equation}
W_0=\frac{7}{9}\frac{\alpha m^2}{\pi \omega_e(n_a)}=
\frac{28}{9}\frac{Z^2\alpha^3}{m^2}n_a L_0.
\label{14}
\end{equation}

The general expression for integral probability of pair creation
by a photon under the simultaneous action of multiple scattering
and an external constant field was obtained in \cite{BK} (see 
Eqs.(2.14) and (1.12)). For further analysis and numerical
calculation it is convenient to turn the contour of integration 
over $t$ at the angle $-i\pi/4$. We obtain after substitution
$t \rightarrow \sqrt{2}t$
\begin{eqnarray}
&& W=\frac{\alpha m^2}{2\pi \omega}\int_0^1 \frac{dy}{y(1-y)}
\int_0^{x_0}\frac{dx}{x_0}G(x, y),\quad G(x, y)=\int_0^{\infty}
F(x, y, t)dt +s_3\frac{\pi}{4},
\nonumber \\
&& F(x, y, t)={\rm Im}\left\lbrace e^{f_1(t)}\left[s_2\nu_0^2
(1+ib)f_2(t)-s_3f_3(t) \right] \right\rbrace,\quad
b=\frac{4\kappa_1^2}{\nu_0^2}, \quad
y=\frac{\varepsilon}{\omega},
\nonumber \\
&& f_1(t)=(i-1)t+b(1+i)(f_2(t)-t),\quad 
f_2(t)=\frac{\sqrt{2}}{\nu_0}\tanh\frac{\nu_0t}{\sqrt{2}},
\nonumber \\
&&f_3(t)=\frac{\sqrt{2}\nu_0}{\sinh(\sqrt{2}\nu_0t)},
\label{15}
\end{eqnarray}
where
\begin{equation}
s_2=y^2+(1-y)^2,~s_3=2y(1-y),~\nu_0^2=4y(1-y)
\frac{\omega}{\omega_c(x)},~\kappa_1=y(1-y)\kappa(x),
\label{16}
\end{equation}
$\varepsilon$ is the energy of one of the particles of pair,
the function $\omega_c(x)$ is defined in Eq.(\ref{11}) and
$\kappa(x)$ is defined in Eq.(\ref{4}).

In order to single out the influence of the multiple 
scattering (the LPM effect) on the process under consideration,
we should consider both the coherent and incoherent contributions.
The probability of coherent pair creation is
(see Eq.(12.7) in \cite{BKS})
\begin{equation}
W^F=\frac{\alpha m^2}{2\sqrt{3}\pi\omega}\int_0^1\frac{dy}{y(1-y)}
\int_0^{x_0}\frac{dx}{x_0}\left[2s_2K_{2/3}(\lambda)
+s_3\int_{\lambda}^{\infty}K_{1/3}(z)dz \right],
\quad \lambda=\frac{2}{3\kappa_1}. 
\label{17}
\end{equation}
The probability of incoherent pair creation is
(compare with Eq.(21.31) in \cite{BKS})
\begin{equation}
W^{inc}=\frac{4Z^2\alpha^3 n_a L}{15 m^2}\int_0^1 dy
\int_0^{\infty}\frac{dx}{\eta_1}e^{-x/\eta_1}f(x, y),
\label{18}
\end{equation}
where $L$ is defined in Eq.(\ref{10}),
\begin{eqnarray}
&&f(x, y)=z^4\Upsilon(z)-3z^2\Upsilon'(z)-z^3+
s_2\left[(z^4+3z)\Upsilon(z)-5z^2\Upsilon'(z)-z^3 \right],
\nonumber \\
&&z=z(x, y)=\kappa_1^{-2/3}. 
\label{19}
\end{eqnarray}
Here 
\begin{equation}
\Upsilon(z)=\int_0^{\infty}\sin\left(zt+\frac{z^3}{3}\right)dt 
\label{20}
\end{equation}
is the Hardy function. For further analysis and numerical 
calculation it is convenient to use the following 
representation of the Hardy function and its derivative
\begin{eqnarray}
&&\Upsilon(z)=\int_0^{\infty}\sin
\left(\frac{\sqrt{3}}{2}z\tau+\frac{\pi}{6}\right)
\exp\left(-\frac{z\tau}{2}-\frac{\tau^3}{3}\right) d\tau 
\nonumber \\
&&\Upsilon'(z)=\int_0^{\infty}\cos
\left(\frac{\sqrt{3}}{2}z\tau+\frac{\pi}{6}\right)
\exp\left(-\frac{z\tau}{2}-\frac{\tau^3}{3}\right)\tau d\tau
\label{21}
\end{eqnarray}

The probabilities $W$ Eq.(\ref{15}), $W^F$ Eq.(\ref{17}), and
$W^{inc}$ Eq.(\ref{18}) at different temperatures T are shown 
in Fig.1 as a function of 
photon energy $\omega$. In low energy region ($\omega \leq 1$~GeV)
one can neglect the coherent process probability $W^F$ as well
as influence of axis field on the incoherent process probability
and the LPM effect and the probability of process is 
$W^{LE}=n_a\sigma_p$  Eq.(\ref{7}).
As one can see in Fig.1 in this energy region the probability
$W$ is by 10\% at T=293 K and  by 20\% at T=100 K less than
the probability at random orientation $W^{ran}$ which is 
taken as $W^{ran}=W^{BM}$ (the Bethe-Maximon probability is 
$W^{BM}=W_0(1-1/42L_0)$=2.17 1/cm in tungsten).

With energy increase the influence of axis field begins and
the LPM effect manifests itself according to Eq.(\ref{13})
(the terms with $\overline{\kappa^2}$ and $(\omega g/\omega_0)^2$
correspondingly). This leads first 
to not large increase of the 
probability $W^{inc}$ which attains the maximum at 
$\omega \sim \omega_m$. The probability $W^F$ in this region
is defined by Eq.(\ref{5a}) and its contribution is relatively
small. The probability $W^F$ becomes comparable with $W^{inc}$
at $\omega \simeq 1.5\omega_m$. At higher energies $W^F$
dominates, while $W^{inc}$ decreases monotonically.

In Fig.2 the calculated total integral probability $W$ of pair 
creation by a photon Eq.(\ref{15}) is compared with data of
NA43 CERN experiment \cite{KKM}. The enhancement is the ratio
$W/W^{BM}$. One can see that the theory
quite satisfactory describes data. This statement differs
from conclusion made in \cite{KKM}. One of reasons for this
difference is diminishing of incoherent contribution 
(see Fig.1): for W, $<111>$, T=100 K at photon energy $\omega=55$~GeV
one has $W^{inc}=0.35W^{BM}$, while in \cite{KKM} 
it was assumed that $W^{inc}=W^{BM}$. 

The contribution of the LPM effect in the total probability 
$W$ Eq.(\ref{15}) is defined as
\begin{equation}
W^{LPM}=W - W^F -W^{inc}
\label{23}
\end{equation}
The relative contribution (negative since the LPM effect suppresses
the process) $\Delta=-W^{LPM}/W$ is shown in Fig.3. This contribution has 
the maximum $\Delta \simeq 5.5$\% at $\omega \simeq 7$~GeV for T=293 K and 
$\Delta \simeq 4.3$\% at $\omega \simeq 12$~GeV for T=100 K or, 
in general, at $\omega \sim\omega_m$. The left part of the curves
is described by the term with $(\omega g/\omega_0)^2$ in
Eq.(\ref{13}). So the rather prevalent assumption that the LPM effect can
essentially suppress the pair creation process in oriented crystals 
is proved wrong due to action of axis field.
On the other hand, the LPM effect can be observed in accurate
measurements. For observation the LPM effect of mentioned scale
in an amorphous tungsten the photons with energy 
$\omega \simeq 10$~TeV are needed \cite{BK1}.

\vspace{0.5cm}

{\bf Acknowledgments}

We are grateful to U.Uggerhoj for discussion of experiment and data.
The authors are indebted to the Russian Foundation for Basic
Research supported in part this research by Grant 
03-02-16154.

\newpage

{\bf Figure captions}

{\bf Fig.1}
Pair creation probability in tungsten, axis $<111>$ at 
different temperatures T.
Curves 1 and 3 are the total probability $W$ Eq.(\ref{15}) for
T=293 K and T=100 K, the curves 2 and 4 give the coherent contribution
$W^F$ Eq.(\ref{17}), the curves 5 and 6 give the incoherent 
contribution $W^{inc}$ Eq.(\ref{18}) at corresponding temperatures T.

{\bf Fig.2}

Enhancement of the probability of pair creation in tungsten,
axis $<111>$. The data are from \cite{KKM}.

{\bf Fig.3}

The relative contribution of the LPM effect $\Delta$ 
(per cent) in tungsten, axis $<111>$.
Curve 1 is for T=293 K and curve 12 is for T=100 K.

\newpage
\begin{table}
\begin{center}
{\sc Table 1}~
{Parameters of the tungsten crystal, axis $<111>$ 
for different temperatures T}
\end{center}
\begin{center}
\begin{tabular}{*{10}{|c}|}
\hline T(K)&$V_0$(eV)&$a_s(10^{-8}$cm)&$x_{0}$&$\eta_1$&
$\omega_0$(GeV)&$\eta$&$\omega_s$(GeV)&$\omega_m$(GeV)&$h$ \\
\hline 293&413&0.215&39.7&0.108&29.7&0.115&34.8&14.35&0.348\\
\hline 100&355&0.227&35.7&0.0401&12.25&0.0313&43.1&8.10&0.612\\
\hline
\end{tabular}
\end{center}
\end{table}


\newpage

\begin{thebibliography}{99}
\bibitem{BKS1} V. N. Baier, V. M. Katkov, and V. M. Strakhovenko,
Sov.Phys.JETP {\bf 63}, 467 (1986).
\bibitem{BKS2} V. N. Baier, V. M. Katkov, and V. M. Strakhovenko,
phys.stat.solidi(b) {\bf 149}, 521 (1988).
\bibitem{BKS} V. N. Baier, V. M. Katkov and V. M. Strakhovenko,
{\em Electromagnetic Processes at High Energies in Oriented
Single Crystals} (World Scientific Publishing Co, Singapore, 1998).
\bibitem{N} H. Nitta, M. Khokonov, Y. Nagata, and S. Onuki,
Phys.Rev.Lett. {\bf 93}, 180407 (2004).
\bibitem{KKM} K. Kirsebom {\it et al}, Nucl. Instrum. Methods
Phys.Res. Sect. B {\bf 135}, 143 (1998).
\bibitem{B}A. Baurichter {\it et al}, Nucl. Instrum. Methods
Phys.Res. Sect. B {\bf 152}, 472 (1999).
\bibitem{BK} V. N. Baier, and V. M. Katkov,
Phys.Lett. A 286, {\bf 299}, 2001.
\bibitem{BK1} V. N. Baier, and V. M. Katkov,
Phys.Rev., {\bf D 62}, 036008 (2000).




\end{thebibliography}
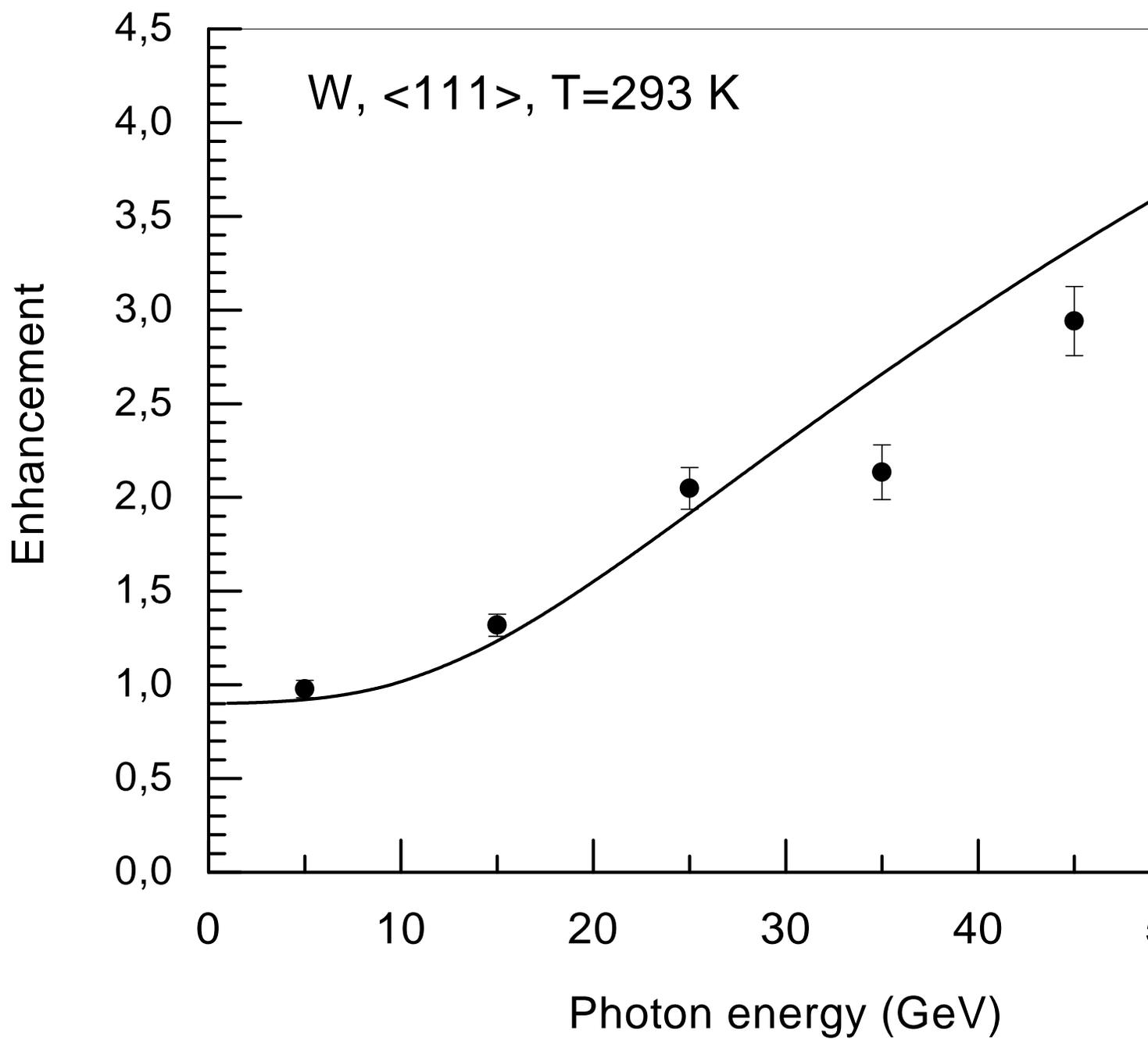
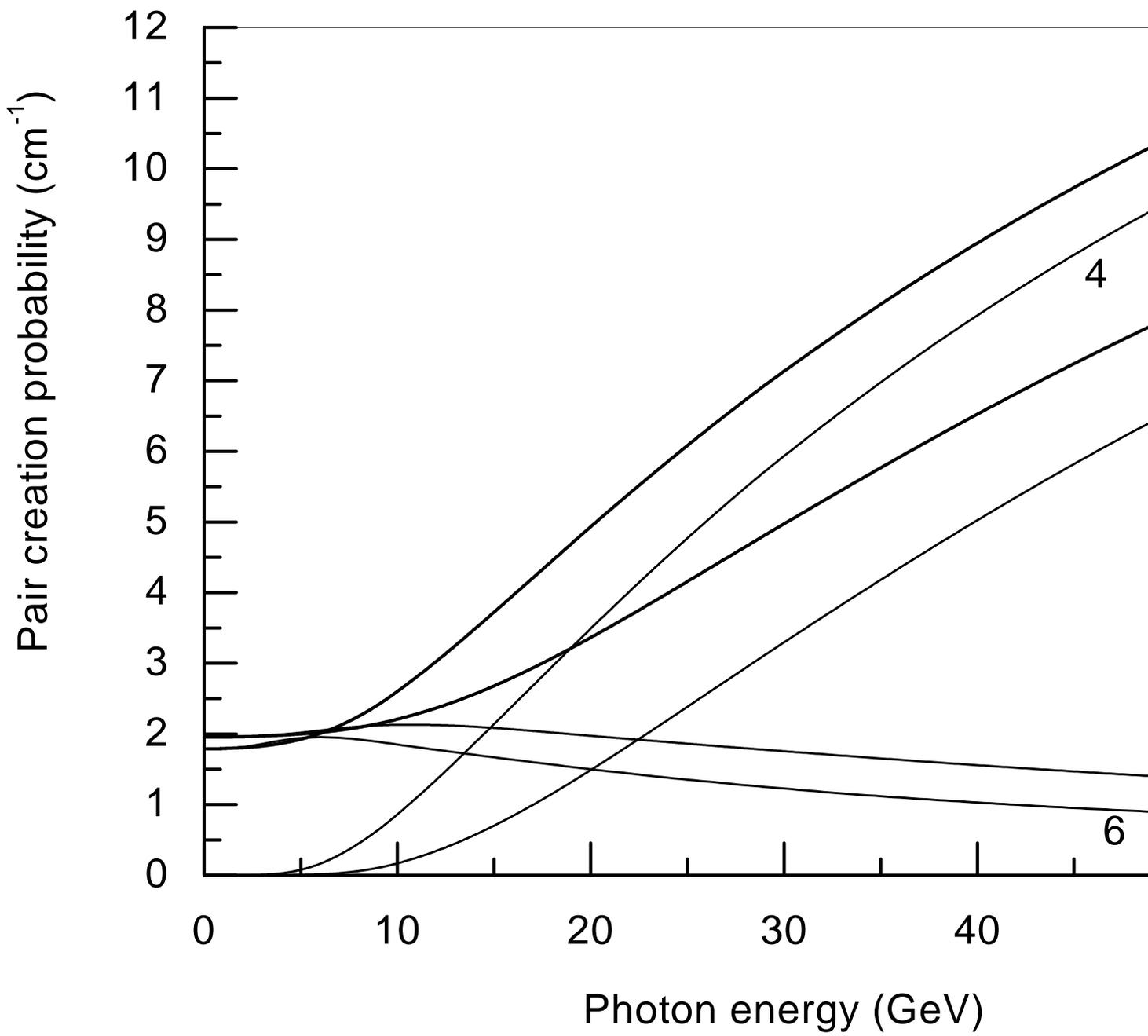
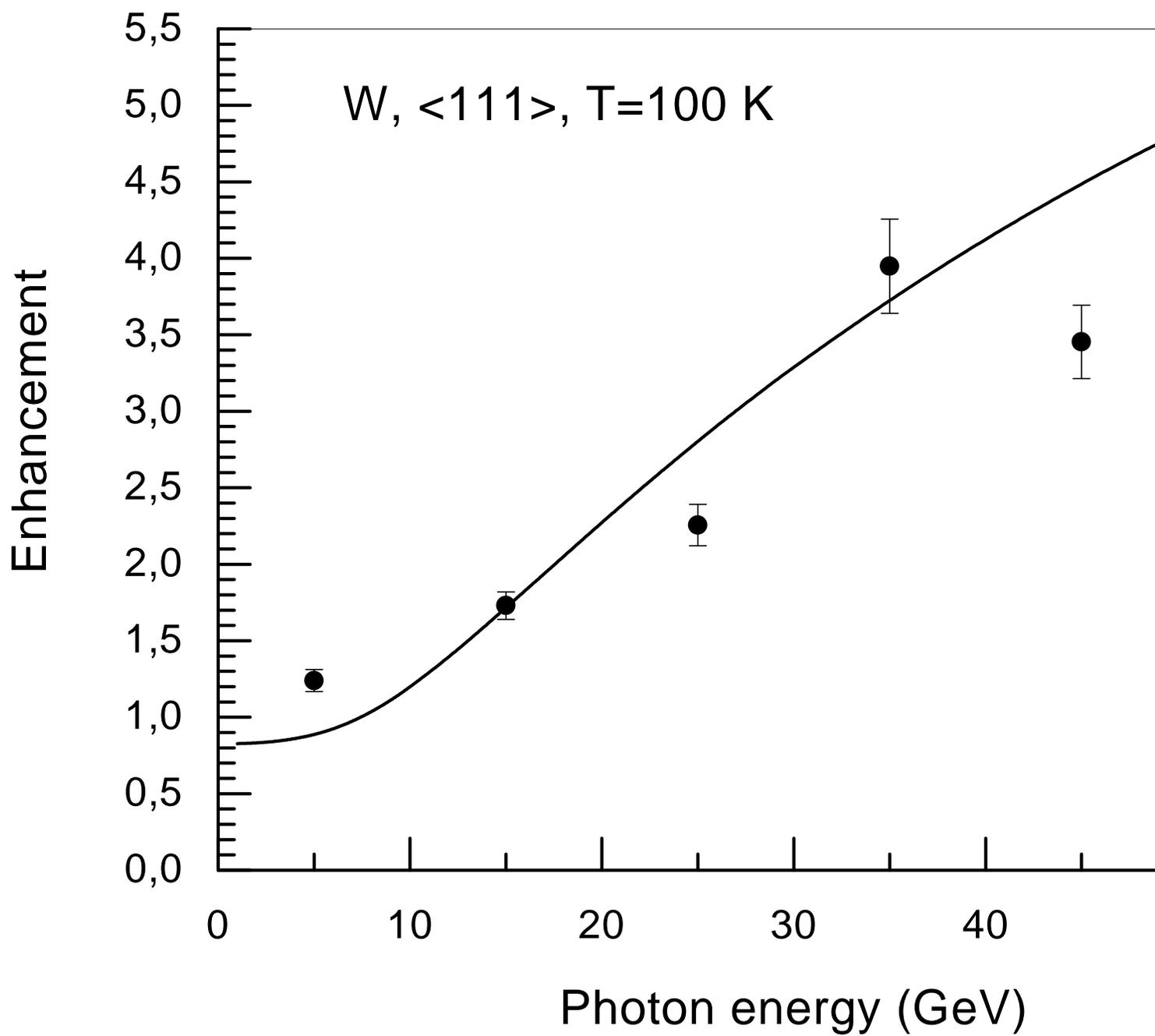
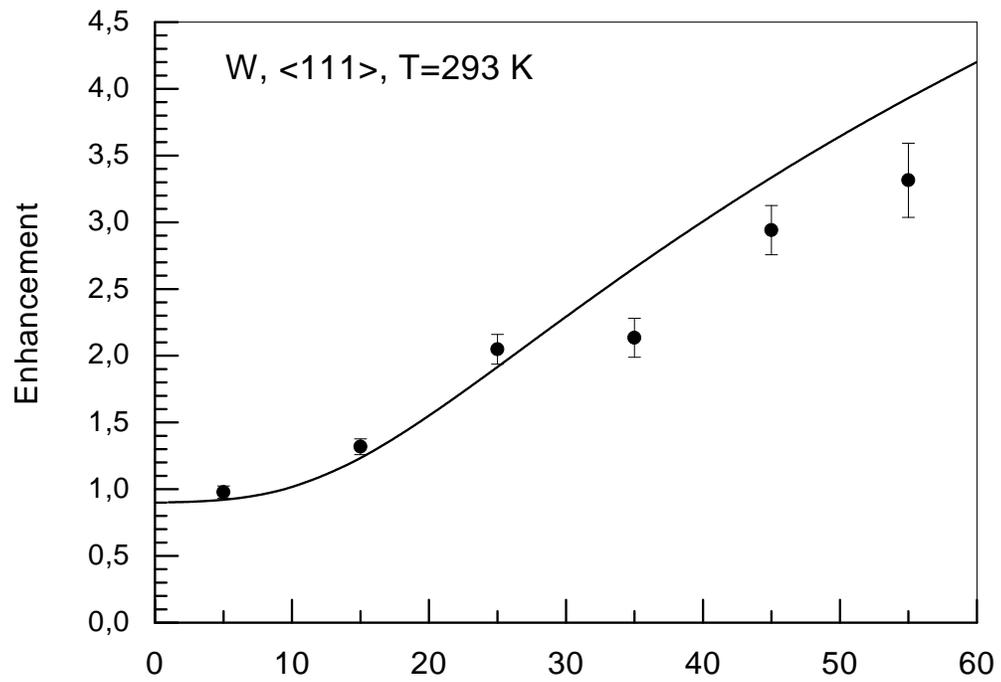
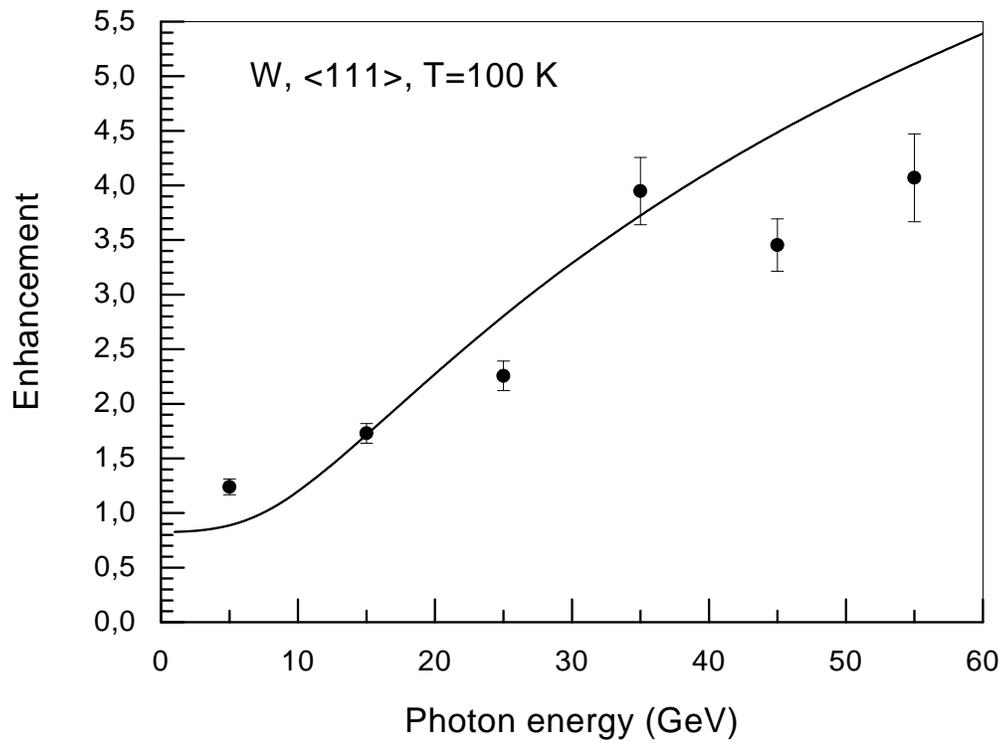
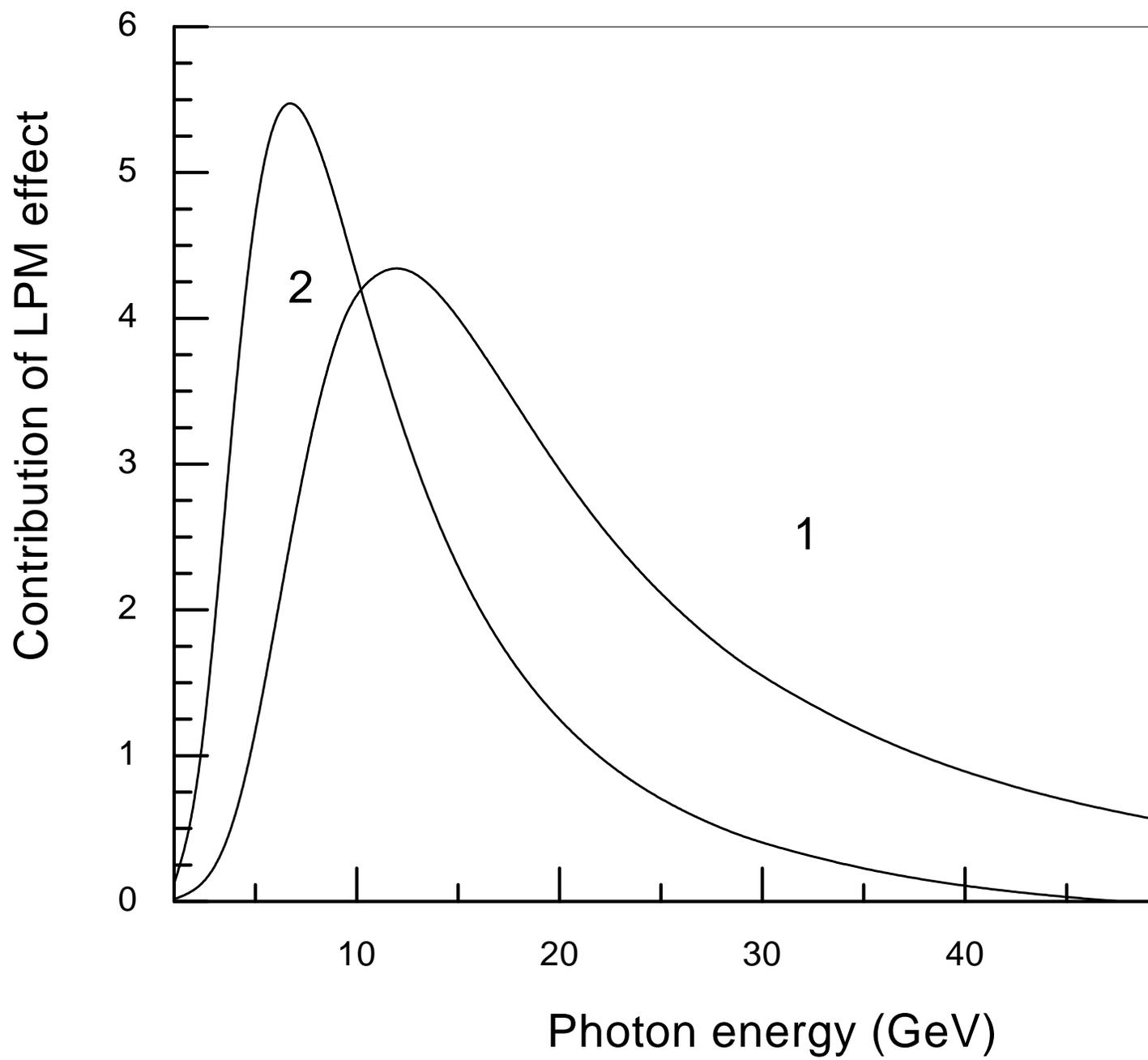
\end{document}